\documentclass[12pt,showpacs,preprintnumbers,superscriptaddress,amsmath,amssymb,nofootinbib]{revtex4}
\usepackage{hyperref}
\usepackage{graphicx}
\usepackage{dcolumn}
\usepackage{bm}
\usepackage{amssymb}
\usepackage{amsmath}
\usepackage{epsfig}    
\usepackage{color}
\usepackage{slashed}
\usepackage{hhline}

\usepackage[normalem]{ulem}
\newcommand\redout{\bgroup\markoverwith
{\textcolor{red}{\rule[.5ex]{2pt}{0.4pt}}}\ULon}
\usepackage{xcolor}

\def\be{\begin{equation}}
\def\ee{\end{equation}}

\newcommand{\bea}{\begin{eqnarray}}
\newcommand{\eea}{\end{eqnarray}}

\newcommand{\nl}{\nonumber\\}

\newcommand{\wt}{\widetilde}
\renewcommand{\ol}{\overline}
\renewcommand{\dag}{\dagger}

\newcommand{\al}{\alpha}
\newcommand{\ga}{\gamma}
\newcommand{\ep}{\epsilon}
\newcommand{\la}{\lambda}
\newcommand{\si}{\sigma}

\numberwithin{equation}{section}

\usepackage{ulem}

\begin{document}
\allowdisplaybreaks[2]

\title{Leptoquark explanation of $h \to \mu\tau$ {and muon $(g-2)$}}

\preprint{KIAS-P15052}

\author{Seungwon Baek}
\email{swbaek@kias.re.kr}
\affiliation{School of Physics, KIAS, 85 Hoegiro, Seoul 02455, Korea}
\author{Kenji Nishiwaki}
\email{nishiken@kias.re.kr}
\affiliation{School of Physics, KIAS, 85 Hoegiro, Seoul 02455, Korea}


\begin{abstract}
We consider lepton flavor violating Higgs decay, specifically $h \to \mu\tau$, in a leptoquark model.
We introduce two {scalar} leptoquarks with the $SU(3)_c \times SU(2)_L \times U(1)_Y$ quantum numbers, 
$(3,2,7/6)$ and $(3,2,1/6)$, which do not generate {dimension-4 operators mediating proton decay.}
They can mix with each other by interactions with the {standard model} Higgs.
The constraint from the charged lepton flavor violating process, $\tau^{-} \to \mu^{-} \gamma$,
is very strong when only one leptoquark contribution is considered.
However, we demonstrate that significant cancellation is possible between the two leptoquark contributions. 
We show that we can explain the CMS (ATLAS) excess in $h \to \mu \tau$.
We also show that muon $(g-2)$ anomaly can also be accommodated.
\end{abstract}
\maketitle
\newpage


\section{Introduction}

Leptoquarks (LQs) are scalar particles which carry both baryon and lepton numbers~\cite{Buchmuller:1986zs}.
They appear in gauge theories with ``unified'' gauge groups, such as Pati-Salam model, $SU(5)$ grand unification, {etc}.

Since LQs are strongly interacting particles which can decay {semileptonically}, their masses are strongly bounded
by the LHC experiments, such as ATLAS and CMS.
For the third-generation scalar LQs, the ATLAS group {excludes the mass in the range}
$m_{\text{LQ}} < 625\,\text{GeV}$ and $200\,\text{GeV} < m_{\text{LQ}} < 640\,\text{GeV}$ at $95\%$ {confidence level}~(C.L.) based on their $8\,\text{TeV}$ data, assuming $100\%$ branching fractions into $b \nu_\tau$ and $t \nu_\tau$, respectively~\cite{Aad:2015caa}.
On the other hand, the CMS group had reported various $8\,\text{TeV}$ bounds at $95\%$ C.L. on $m_{\text{LQ}}$ as 
$m_{\text{LQ}} \,{>}\, 740\,\text{GeV}$, $m_{\text{LQ}} \,{>}\, 650\,\text{GeV}$ and $m_{\text{LQ}} \,{>}\, 685\,\text{GeV}$ with assumptions of $100\%$ branching fractions into $b\tau$, $t\nu_\tau$ and $t \tau$, respectively~\cite{Khachatryan:2014ura,Khachatryan:2015bsa}.

We note that the CMS excess of $e e j j$ and $e\nu j j$~\cite{CMSeejj} can also be interpreted as a signal of the first generation LQ with mass
about 650 GeV.
An example of detailed study of LQ models for the excess can be found in~\cite{Queiroz:2014pra}.

In the {standard model}~(SM), {lepton flavor violating (LFV) Higgs decay channels} are absent at tree level and highly suppressed by
small neutrino masses and {the} GIM mechanism at loop level. 
Therefore, once they are observed with sizable branching fractions, they indicate {a} clear signal of new physics beyond the SM.
The CMS collaboration reported the LFV Higgs decay branching fraction, using the 19.7 fb$^{-1}$ of $\sqrt{s} = 8$ TeV,
\bea
 \mathcal{B}(h \to \mu \tau) = {(0.84^{+ 0.39}_{-0.37})} \%,
\label{eq:CMS_h}
\eea
which deviates $2.4 \sigma$ from zero~\cite{Khachatryan:2015kon}.
Here, $\mu\tau$ means the inclusive final state consisting of $\mu^{+} \tau^{-}$ and $\mu^{-} \tau^{+}$.
Although recent ATLAS measurement, using the 20.3 fb$^{-1}$ of $\sqrt{s} = 8$ TeV,
\bea
 \mathcal{B}(h \to \mu \tau) = (0.77 \pm 0.62) \%,
\eea
does not show {a} significant deviation from the SM~\cite{Aad:2015gha}, it is at least consistent with the CMS result.
If confirmed by the future data at LHC Run II which can probe down to $\sim 10^{-3}$, it would be a clear signal requiring new physics beyond the
SM. There are several model-independent~\cite{Harnik:2012pb,Blankenburg:2012ex,Davidson:2012ds,Kopp:2014rva,Celis:2014roa,Lee:2014rba,deLima:2015pqa,Varzielas:2015iva,Bhattacherjee:2015sia,Mao:2015hwa,He:2015rqa,Goto:2015iha,Altmannshofer:2015esa}
and also model-dependent studies~\cite{Pilaftsis:1992st,Arhrib:2012mg,Falkowski:2013jya,Dery:2014kxa,Campos:2014zaa,Sierra:2014nqa,
Heeck:2014qea,Crivellin:2015mga,Dorsner:2015mja,Omura:2015nja,Crivellin:2015lwa,Das:2015zwa,
Chiang:2015cba,Crivellin:2015hha,
Cheung:2015yga,Botella:2015hoa,Baek:2015fma}
to accommodate this deviation.

The theoretical and experimental sensitivity of {the} anomalous magnetic moment of {the} muon, {\it i.e.} $(g-2)_\mu$, has reached
to probe the electroweak scale.
State of the art calculations in the SM cannot explain the experimental result, and there is
about 3$\sigma$ discrepancies between them~\cite{Jegerlehner:2009ry,Benayoun:2012wc}:
\bea
 \Delta a_\mu = a_\mu^{\rm exp} - a_\mu^{\rm SM} = (299 \pm 90 ~~\text{to}~~ 394 \pm 84) \times 10^{-11},
\eea
which also calls for new physics models.

In this paper we consider a LQ model as an explanation of the LFV Higgs decay, $h \to \mu\tau$ {and} muon $(g-2)$
anomaly.
Considering the proton decay constraints, only two {types of $SU(2)_L$} doublet leptoquarks are favored.
We assume both of them are realized in nature. We first show that a strong constraint from $\tau^{-} \to \mu^{-} \gamma$ 
can be alleviated significantly due to cancellations between the top and bottom quark contributions.\footnote{Note that similar discussions in the context of LQs are found in~\cite{Dorsner:2015mja,Cheung:2015yga}.}
We show that there is allowed parameter space to accommodate the $h \to \mu\tau$ anomaly {and} $(g-2)_\mu$.
{A smoking gun signal which distinguishes our model from other models would be the direct LQ production at colliders.
A promising signature at the LHC is the pair production of LQs decaying into a quark and a lepton, where the decay pattern is so characteristic.
Especially, components with $+2/3$ electric charge, named as $Y_1$ and $Y_2$ later, should be relatively light and couple to the bottom quark, the electron and the muon for the explanation of the excess in $h \to \mu\tau$ and the muon $(g-2)$ with circumventing the bound from $\tau^- \to \mu^- \gamma$.
Therefore, the smoking gun final states in our model are $b \ol{b} \, \tau^- \tau^+$ and $b \ol{b} \, \mu^- \mu^+$.}

The paper is organized as follows.
In Sec.~\ref{sec:model}, we introduce our model.
In Sec.~\ref{sec:tau2mug}, we consider $\tau^{-} \to \mu^{-} \gamma$ constraint.
In Sec.~\ref{sec:h2mt}, we consider $h \to \mu\tau$ signal.
In Sec.~\ref{sec:g-2}, we show that we show that we can accommodate $(g-2)_\mu$.
In Sec.~\ref{sec:conc}, we summarize and conclude.

\section{The Model}
\label{sec:model}

Among the possible  LQs which have renormalizable interactions with the SM
fermions, only $R_2$ and $\wt{R}_2$ in the notation of~\cite{Buchmuller:1986zs} do not have problem with the constraint from the proton decay {within renormalizable perturbation theory}~\cite{Arnold:2013cva,Queiroz:2014pra}.
They are in the representation 
\bea
R_2(3, 2, 7/6), \quad  \wt{R}_2(3, 2, 1/6), 
\eea
in the SM gauge group $SU(3)_C \times SU(2)_L \times U(1)_Y$.\footnote{{In the case of non-$SU(2)_L$-doublet LQs, we can write down gauge-invariant dimension-four operators generating rapid proton decay.
The $SU(2)_L$ doublet ones {do not allow such dangerous operators at renormalizable level.}
However, as discussed in~\cite{Arnold:2013cva,Queiroz:2014pra}, constraints from dimension-five effective operators (generating proton decay) are still severe, where $m_{\text{LQ}}$ should be greater than around $10^{5}\,\text{TeV}$ even when the cutoff scale is equal to the Planck scale.
A remedy for reducing $m_{\text{LQ}}$ is to introduce a new symmetry prohibiting the operators.
In this paper, we do not consider constraints from the proton decay caused by higher dimensional operators.}
}

Assuming both of them exist in nature at renormalizable level, they interact with quarks and leptons via the interaction Lagrangian 
\bea
{\cal L} &=& -\la_u^{ij} \ol{u}_R^i R_2^T \ep \, L_L^j -\la_e^{ij} \ol{e}_R^i R_2^\dag Q_L^j -\la_d^{ij} \ol{d}_R^i \wt{R}_2^T \ep \, L_L^j  +{ h.c.},
\eea
where we have suppressed color indices and $\ep \,(\equiv i\sigma^2)$ is the two-by-two {antisymmetric} matrix with $\ep^{12}=1$.
The scalar potential is given by
\bea
V &=& \mu_H^2 |H|^2 + \mu_2^2 |R_2|^2  + \wt{\mu}_2^2 |\wt{R}_2|^2 \nl
&+&  \la_H |H|^4 + \la_2 |R_2|^4 + \wt{\la}_2 |\wt{R}_2|^4 + \la_{HR} |H|^2 |R_2|^2 + \wt{\la}_{HR} |H|^2 |\wt{R}_2|^2 \nl
&+&\la_{H2} R_2^\dagger H H^\dagger R_2  +\wt{\la}_{H2} \wt{R}_2^\dagger H H^\dagger {\wt{R}_2}  
+ \left(\la_{\rm mix} R_2^\dag H \wt{R}_2 \ep \, H + h.c.\right),
\eea
where $H(1,2,1/2)$ is the SM Higgs doublet.
{$R_2$ and $\wt{R}_2$ fields} can be decomposed into {$SU(2)_L$} components,
\bea
R_2 =\left(\begin{array}{c} V \\ Y \end{array}\right), \quad
\wt{R}_2 =\left(\begin{array}{c} \wt{Y} \\ \wt{Z} \end{array}\right). 
\eea
After the Higgs gets vacuum expectation value (vev), {$v\,(\simeq 246\,\text{GeV})$}, we can write
\bea
H=\left(\begin{array}{c} 0 \\ \frac{1}{\sqrt{2}}(v+h) \end{array}\right), 
\eea
in the unitary gauge. Then, the masses of $V$ and $\wt{Z}$ are given by
\bea
m_V^2 = \mu_2^2 + {1 \over 2} \la_{HR} v^2,\quad
 m_{\wt{Z}}^2 = \wt{\mu}_2^2 +\frac{1}{2} \wt{\la}_{HR} v^2 +\frac{1}{2} \wt{\la}_{H2} v^2.
\eea
The mass terms of $Y$ and $\wt{Y}$ are written as
\bea
{\cal L}_{\rm mass}(Y,\wt{Y})=-\left(\begin{array}{cc} Y^\dag & \wt{Y}^\dag \end{array}\right) 
\left(\begin{array}{cc} \mu_2^2 +\frac{1}{2} \la_{HR} v^2 +\frac{1}{2} \la_{H2} v^2 &  \frac{1}{2} \la_{\rm mix} v^2 \\
                                  \frac{1}{2} \la_{\rm mix} v^2 & \wt{\mu}_2^2 +\frac{1}{2} \wt{\la}_{HR} v^2\end{array}\right) 
\left(\begin{array}{cc} Y \\ \wt{Y} \end{array}\right).
	\label{eq:LQ_mass_matrix}
\eea
The mass eigenstates, $Y_1, Y_2$ {(with the electromagnetic charge $+2/3$)} are mixture of $Y$ and $\wt{Y}$ with mixing angle $\alpha_Y$,
\bea
\left(\begin{array}{cc} Y \\ \wt{Y} \end{array}\right) =
\left(\begin{array}{cc} c_Y & s_Y\\
                                -s_Y & c_Y\end{array}\right)
\left(\begin{array}{cc} Y_1 \\ Y_2 \end{array}\right) 
\equiv O
\left(\begin{array}{cc} Y_1 \\ Y_2 \end{array}\right) , 
\label{eq:mixingY}
\eea
{where} $c_Y=\cos\al_Y, s_Y=\sin\al_Y$.

As we will see, large $\alpha_Y$ and large mass splitting between $V$ and {$Y_i$} are favored to satisfy the experimental constraints and also to enhance $h \to \mu \tau$.
Concretely speaking, the relation $m_{Y_{i}} \sim m_{V}/6$ will be {imposed to avoid} the bound from $\tau^{-} \to \mu^{-} \gamma$ {naturally} with sizable couplings, which are required for explanations of the excess in $h \to \mu \tau$.

Here, we look into the mass matrix in Eq.~(\ref{eq:LQ_mass_matrix}) and discuss whether we can realize the mass hierarchy as $m_{Y_{i}} \sim m_{V}/6$ with a large mixing in $\alpha_Y$ in our setup.
A key point is that the $(1,1)$ component of the mass matrix is rephrased as $m_V^2 + \frac{1}{2} \lambda_{H2} v^2$.
Then, when the following relations are realized,
\bea
m_V^2,\, \lambda_{H2} v^2 \ > \ \lambda_{\text{mix}} v^2,\, \wt{\mu}_2^2 +\frac{1}{2} \wt{\la}_{HR} v^2,
\eea
and a cancellation {occurs} between $m_V^2$ and $\frac{1}{2} \lambda_{H2} v^2$ with a negative $\lambda_{H2}$, the relation $m_{Y_{i}} \sim m_{V}/6\,(i=1,2)$ can be realized.
In addition, if the off-diagonal terms are comparable with diagonal ones, a large mixing angle in $\alpha_Y$ is expected.
For example, if {the} (1,1) and (2,2) components
are of similar size with $\sim {\cal O}(1)$  TeV$^2$ and $\lambda_{\rm mix} \sim {10}$, we get ${m_{Y_{1,2}} \gtrsim} ~ 0.7~ {\rm TeV}$ and {a}
maximal $\alpha_Y$, which can obviously avoid the current direct search bound on third generation LQs.
However, when $m_V$ is multi TeV, a realization of {such} cancellation between $m_V^2$ and $\frac{1}{2} \lambda_{H2} v^2$ would get to be nontrivial within perturbative $\lambda_{H2}$.
To further enhance mass difference between $V$ and {$Y_i$} and/or the mixing angle $\alpha_Y$, we can implicitly assume {additional contributions via} higher dimensional operators such as
\bea
\frac{\phi}{\Lambda} R_2^\dagger H H^\dagger R_2,\quad
\frac{\phi}{\Lambda} R_2^\dag H \wt{R}_2 \ep \, H,
\label{eq:dim5}
\eea
where $\phi$ is a new singlet with {a large vev as $\langle \phi \rangle > \Lambda$}.

{Although there is no apparent symmetry which leads to the mass ratio $m_{Y_{i}} \sim m_{V}/6$, the UV
complete {grand unified theory~(GUT)} or flavor theory into which our low energy effective theory is embedded have larger symmetry and we
expect they guarantee the mass ratio without fine tuning. The high energy theory will also generate the dimension-five operators in (\ref{eq:dim5}).}

Finally, we comment on the case with a small mixing angle $\alpha_Y$, which corresponds to the possibility that no sizable cancellation {occurs} between the terms $m_V^2$ and $\frac{1}{2} \lambda_{H2} v^2$.
Even in this case, (at least) one mass eigenstate can be light as the relation $m_{Y_{i}} \sim m_{V}/6$ being fulfilled.
But, as we will see in {Sec.}~\ref{sec:tau2mug}, we should accept a larger hierarchy between two leptoquark couplings to circumvent the bound from $\tau^{-} \to \mu^{-} \gamma$.
To make matters worse, as discussed in {Sec.}~\ref{sec:h2mt}, such hierarchical couplings {are} inappropriate for explaining the excess in $h \to \mu \tau$.

\section{$\tau^{-} \to \mu^{-} \gamma$}
\label{sec:tau2mug}

In this section, we consider the constraints from the {charged} lepton flavor violating processes.
Since we are interested in {$2 \leftrightarrow 3$} transitions, we restrict ourselves only to  $\tau^{-} \to \mu^{-} \gamma$ decay.
Our study can be applied to other LFVs, such as $\mu^{-} \to e^{-}$ or $\tau^{-} \to e^{-}$ transitions, similarly.
However, we assume they are sufficiently suppressed by small LFV couplings.

\begin{figure}
\center
\includegraphics[width=7cm]{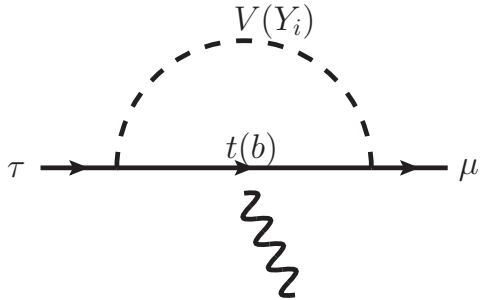}
\caption{Feynman diagrams for $\tau^{-} \to \mu^{-} \gamma$.  The photon line can be attached to any charged particles, and there
are four possibilities.}
\label{fig:tau2mug}
\end{figure}

The effective Hamiltonian for $\tau^{-} \to \mu^{-} \gamma$ is written as
\bea
 {\cal H}_{\rm eff} = C_R^\ga  ~\ol{\mu}_L \si^{\mu\nu} F_{\mu\nu} \tau_R+ C_L^\ga~ \ol{\mu}_R \si^{\mu\nu} F_{\mu\nu} \tau_L,
\label{eq:Heff4t2mg}
\eea
where $C_{R,L}^\ga$ are Wilson coefficients and $F_{\mu\nu}\,(=\partial_\mu A_\nu - \partial_\nu A_\mu)$ is {the} photon field strength tensor.

The Feynman diagrams for $\tau^{-} \to \mu^{-} \gamma$ are shown in Fig.~\ref{fig:tau2mug}.
We note that in our model, the chirality flip appearing in (\ref{eq:Heff4t2mg}) can occur inside the loop.
Therefore the amplitudes can be proportional to $m_t$ or $m_b$ instead of small masses {from the} external lines, $m_\tau$ or $m_\mu$.
This is the main reason that this LFV process becomes a very strong constraint {in ordinary third generation LQ models}. 

The Wilson coefficients $C_{R,L}^\ga$ can be calculated from the diagrams in Fig.~\ref{fig:tau2mug}:
\bea
C_R^\ga &=& \frac{N_c e}{32 \pi^2 m_V^2}\Bigg[ 
\Big(\la_e^{23} \la_e^{33*} m_\mu  +\la_u^{32*} \la_u^{33} m_\tau  \Big)
\Big(\frac{2}{3} I_1(x) +\frac{5}{3} J_1(x) \Big) 
+ \la_u^{32*}  \la_e^{33*}  m_t \Big(\frac{2}{3} I_2(x) +\frac{5}{3} J_2(x) \Big) \Bigg] \nl
&+& \sum_{j=1,2} \frac{N_c e}{32 \pi^2 m_{Y_j}^2}\Bigg[ 
\Big(\la_e^{23} \la_e^{33*} O_{1j}^2 m_\mu  +\la_d^{32*} \la_d^{33} O_{2j}^2 m_\tau  \Big)
\Big(-\frac{1}{3} I_1(y_j) +\frac{2}{3} J_1(y_j) \Big) \nl
&+& \la_d^{32*}  \la_e^{33*}  O_{1j} O_{2j} m_b \Big(-\frac{1}{3} I_2(y_j) +\frac{2}{3} J_2(y_j) \Big) \Bigg], \nl
C_L^\ga &=& \frac{N_c e}{32 \pi^2 m_V^2}\Bigg[ 
\Big(\la_u^{32*} \la_u^{33}  m_\mu  +  \la_e^{23} \la_e^{33*} m_\tau \Big)
\Big(\frac{2}{3} I_1(x) +\frac{5}{3} J_1(x) \Big) 
+ \la_e^{23}  \la_u^{33}  m_t \Big(\frac{2}{3} I_2(x) +\frac{5}{3} J_2(x) \Big) \Bigg] \nl
&+& \sum_{j=1,2} \frac{N_c e}{32 \pi^2 m_{Y_j}^2}\Bigg[ 
\Big(\la_d^{32*} \la_d^{33} O_{2j}^2  m_\mu  + \la_e^{23} \la_e^{33*} O_{1j}^2 m_\tau  \Big)
\Big(-\frac{1}{3} I_1(y_j) +\frac{2}{3} J_1(y_j) \Big) \nl
&+& \la_e^{23}  \la_d^{33}  O_{1j} O_{2j} m_b \Big(-\frac{1}{3} I_2(y_j) +\frac{2}{3} J_2(y_j) \Big) \Bigg], 
\label{eq:C7g}
\eea
where $N_c=3$ is the color factor, $x=m_t^2/m_V^2$, and $y_i = m_b^2/m_{Y_i}^2$.
The loop functions are obtained to be
\bea
I_1(x) &=& \frac{2+3x-6x^2+x^3+6x \log x}{12(1-x)^4}, \nl
J_1(x) &=& \frac{1-6x+3 x^2+2 x^3-6 x^2 \log x}{12(1-x)^4}, \nl
I_2(x) &=& \frac{-3 + 4 x - x^2 -2 \log x}{2(1-x)^3}, \nl
J_2(x) &=& \frac{1 -  x^2  +2 x \log x}{2(1-x)^3}.
\label{eq:loop_ftns}
\eea
The branching ratio of $\tau^{-} \to \mu^{-} \gamma$ is then
\bea
{\cal B}(\tau^{-} \to \mu^{-} \gamma) &=&\frac{\tau_\tau (m_\tau^2-m_\mu^2)^3}{4 \pi m_\tau^3} \left(|C_R^\gamma|^2 +|C_L^\gamma|^2\right),
\eea
where $\tau_\tau=87.03$ $\rm \mu m$ is the lifetime of $\tau$. The current experimental bound is~\cite{BaBar:tau2mug}
\bea
{\cal B}(\tau^{-} \to \mu^{-} \gamma) < 4.4 \times 10^{-8}.
\label{eq:tau2mug_bound}
\eea
This corresponds to
\bea
|C_R^\gamma|^2 +|C_L^\gamma|^2 < \left(4.75 \times 10^{-10} \over {\rm GeV}\right)^2.
\label{eq:CRbound}
\eea

\begin{figure}
\center
\includegraphics[width=0.45\columnwidth]{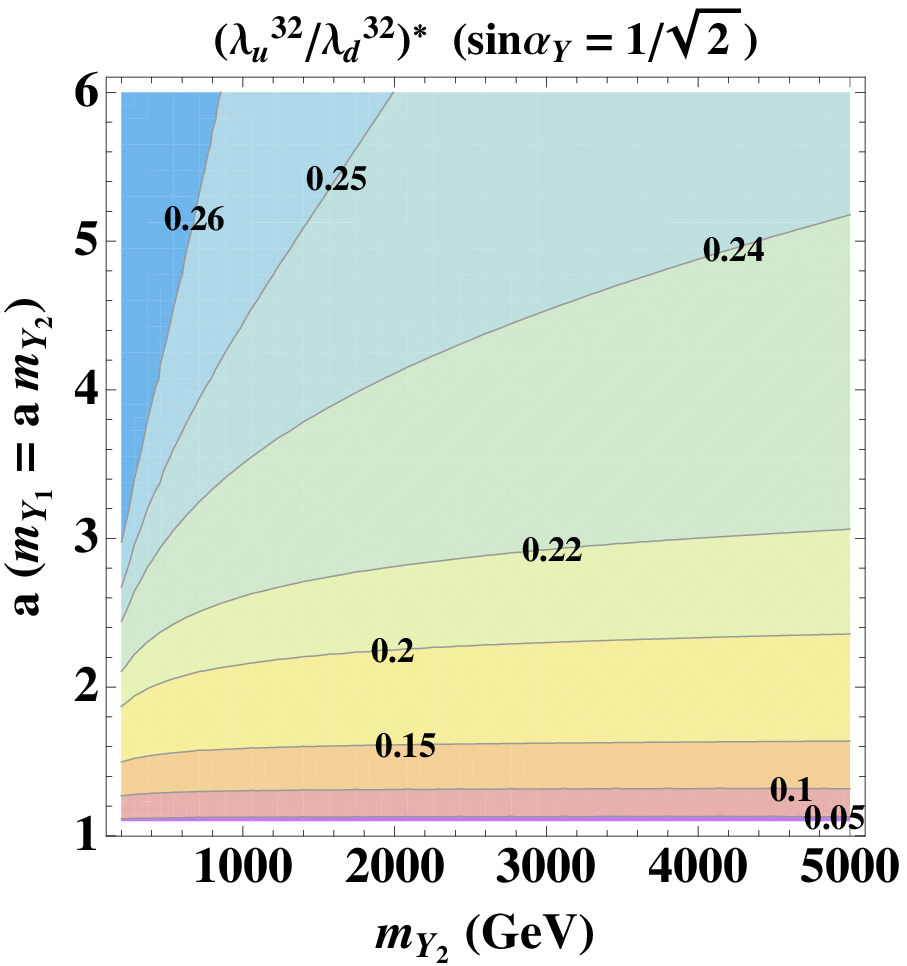}
\includegraphics[width=0.45\columnwidth]{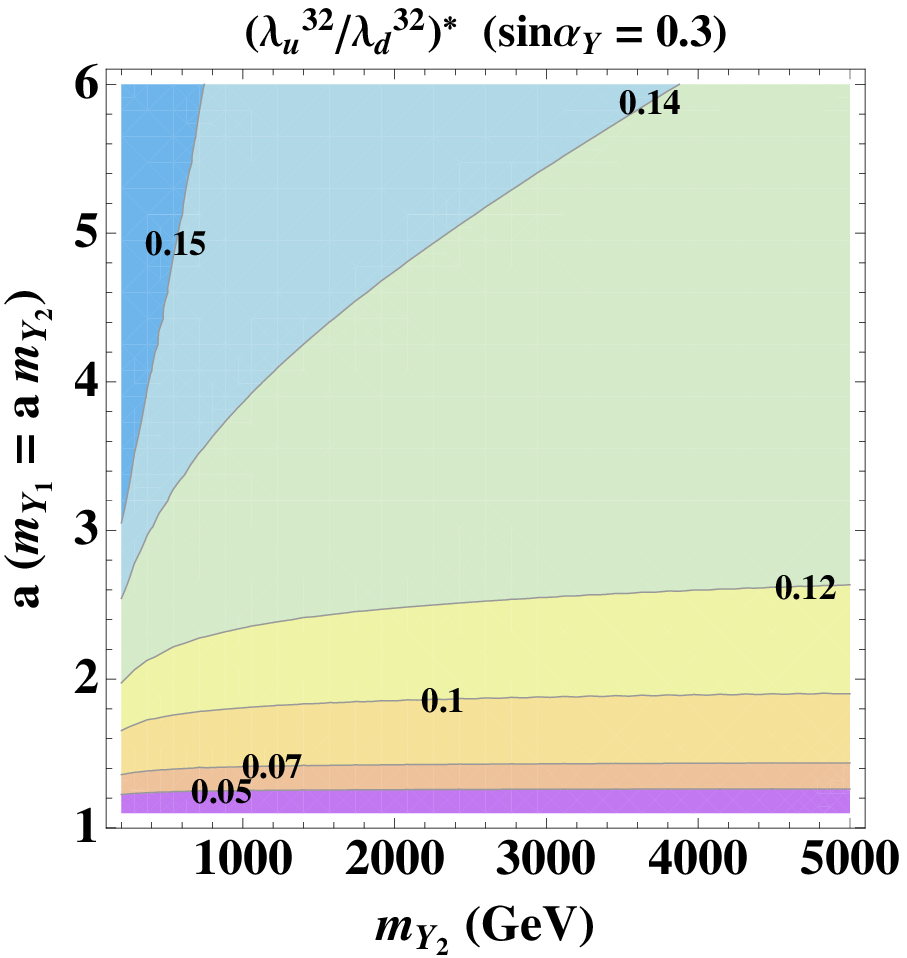}
\includegraphics[width=0.45\columnwidth]{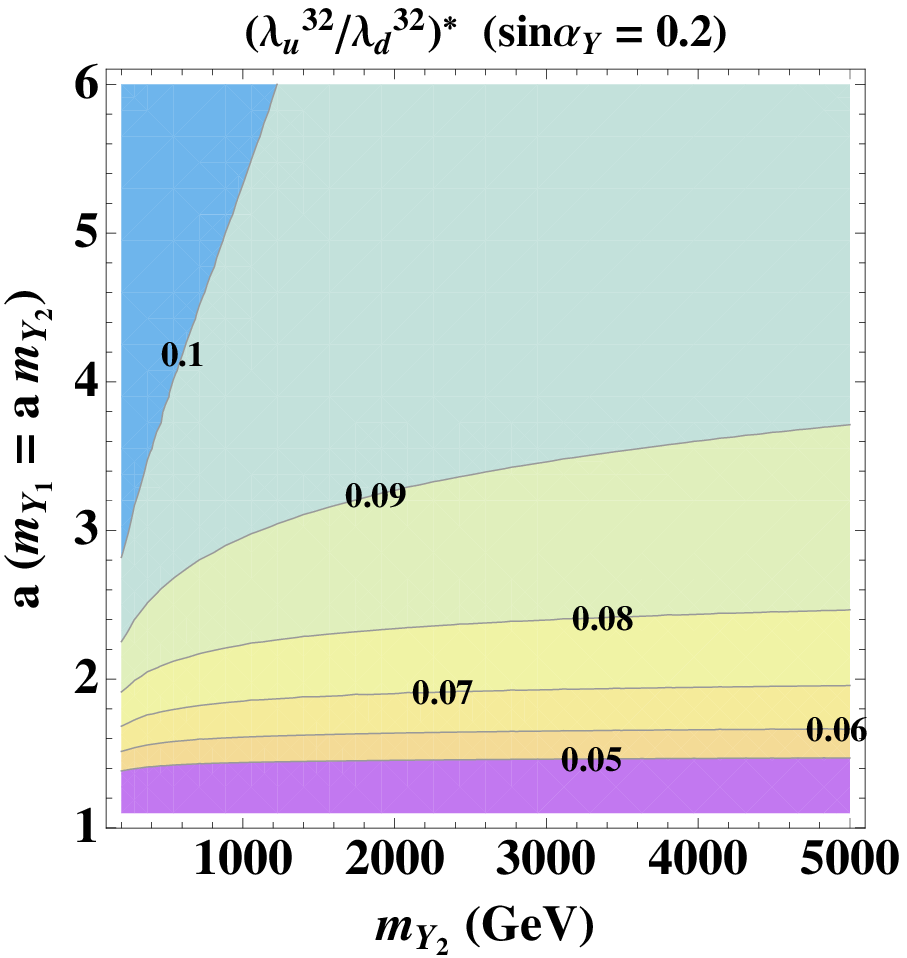}
\caption{{Contour plots for $(\lambda_u^{32}/\lambda_d^{32})^{\ast}$ which is required for {exact} cancellations of $\tau^{-} \to \mu^{-} \gamma$ in {$(m_{Y_2}, m_{Y_1})$}-plane through Eq.~(\ref{eq:exact_cancel}).
From top-left to bottom, the sine of the mixing angle $\sin{\alpha_Y}$ is chosen as $1/\sqrt{2}$, $0.3$ and $0.2$, respectively.
$m_V$ is set as $m_V = 6 \, m_{Y_2}$ and $m_{Y_1}$ is formulated as $m_{Y_1} = a \, m_{Y_2}$ by use of the factor $a$, where the range $[1.1, 6.0]$ is considered in the three plots.}
}
\label{fig:cancel}
\end{figure}

For the discussion of $\tau^- \to \mu^- \gamma$, we assume $C_L^\gamma =0$ for simplicity.
If we consider {a} single leptoquark contribution from $V$, ${\cal B}(\tau^- \to \mu^- \gamma)$ gives lower mass bound $0.85, 3.7, 14, 42$ TeV, 
for $\lambda_e^{33}=0.001, 0.01, 0.1, 1$, respectively, where we took $\lambda_u^{32} =0.35$.
With these parameters, we obtain too small contribution to $h \to \mu \tau$ as was
noticed in~\cite{Dorsner:2015mja}.

Since we introduce both $R_2$ and $\wt{R}_2$, we can have diagrams with {the} chirality flip inside
the $b$-quark loop, which generate terms proportional to $m_b$.
The $Y_i-b$ contributions are naively expected to be smaller than $V-t$ contribution by factor $m_b/m_t \sim 1/35$. However,
since $C_{R,L}^\ga$ are proportional to $m_f/m_{\rm LQ}^2$ ($f=t,b$) as can be seen in (\ref{eq:C7g}),
if $m_{Y_i} \sim  m_V/6$, cancellations between $t$ and $b$ contributions can occur naturally.
Note that a nonzero mixing between $Y$ and $\wt{Y}$ is mandatory for a natural cancellation since in the limit $\sin{\alpha_Y} \to 0$ the contributions being proportional to $m_b$ turn out to be zero.
Neglecting {small} terms proportional to $m_\tau$ or $m_\mu$, an exact cancellation in $C_R^\gamma$ occurs when the following condition is held,
\bea
\frac{\la_u^{32*}}{\la_d^{32*}} = {- \sum_{i=1}^{2} O_{1i} O_{2i} \, \frac{m_b}{m_t} \left(\frac{m_V}{m_{Y_i}}\right)^2
\frac{-\frac{1}{3} I_2(y_i) +\frac{2}{3} J_2(y_i) }{+\frac{2}{3} I_2(x) +\frac{5}{3} J_2(x)}.}
\label{eq:exact_cancel}
\eea
{In Fig.~\ref{fig:cancel}}, the values of the ratio $(\lambda_u^{32}/\lambda_d^{32})^{\ast}$ which are required for {exact} cancellations of $\tau^{-} \to \mu^{-} \gamma$ {are shown as {$(m_{Y_2}, m_{Y_1})$}-planes} through Eq.~(\ref{eq:exact_cancel}) with the three choices of the sine of the mixing angle $\sin{\alpha_Y}$ as $1/\sqrt{2}$, $0.3$ and $0.2$.\footnote{{Note that the sign of $\sin{\alpha_Y}$ is not important. We can compensate a negative sign by flipping the sign of the coupling $\lambda_u^{32}$ or $\lambda_d^{32}$.}}
Here, $m_V$ is set as $m_V = 6 \, m_{Y_2}$ and $m_{Y_1}$ is formulated as $m_{Y_1} = a \, m_{Y_2}$ by use of the factor $a$, where the range $[1.1, 6.0]$ is considered in the three plots.
Note that in the case that $m_{Y_1}$ and $m_{Y_2}$ are completely degenerated, the two contributions being proportional to $m_b$ are exactly {canceled} out between them and no cancellation mechanism works in $\tau^{-} \to \mu^{-} \gamma$.
Here, almost all the shown regions in Fig.~\ref{fig:cancel} (where $\sin{\alpha_Y}$ is greater than $0.2$), the target values of the ratio 
$(\lambda_u^{32}/\lambda_d^{32})^{\ast}$ {are} greater than $0.05$, which means that {we can adjust naturally}  the two couplings for realizing the cancellation.
However, as we will see in the following {section}, when the ratio $(\lambda_u^{32}/\lambda_d^{32})^{\ast}$ gets to be small, it is hard to explain the excess of $h \to \mu\tau$.

\begin{figure}
\center
\includegraphics[width=8cm]{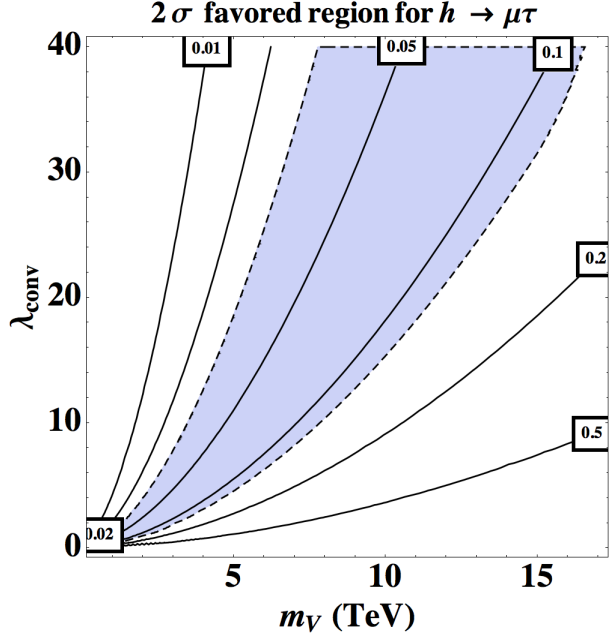}
\caption{The blue region shows the {2$\sigma$} favored region for $h \to \mu\tau$ in $(m_V, \lambda_{\rm conv})$-plane, where
$\lambda_{\rm conv} \equiv |\lambda_u^{32} \lambda_e^{33}|$.
The black contours indicate degrees of the fine tuning defined around Eq.~(\ref{eq:degree_fine_tuning}) {in percentage terms}.}
\label{fig:h2mt_V}
\end{figure}

{The Wilson coefficient $C_{R}^\ga$ can be {rewritten} in terms of the ratio in (\ref{eq:exact_cancel}), which we will
  define as $({\la_u^{32*}}/{\la_d^{32*}})_{\rm cancel}$,
\bea
C_R^\ga \simeq
\frac{N_c e}{32 \pi^2 m_V^2}
 \la_u^{32*}  \la_e^{33*}  m_t \Big(\frac{2}{3} I_2(x) +\frac{5}{3} J_2(x) \Big)
\Bigg[  1 -\frac {\la_d^{32*}}{\la_u^{32*}} \left(\frac{\la_u^{32*}}{\la_d^{32*}}\right)_{\rm cancel} \Bigg].
\label{eq:degree_fine_tuning}
\eea
This equation shows again that, if ${\la_d^{32*}}/{\la_u^{32*}} =({\la_d^{32*}}/{\la_u^{32*}})_{\rm cancel}$,
$C_R^\ga =0$. We can consider a deviation from the exact cancellation by introducing $\delta$ in such a way that
${\la_d^{32*}}/{\la_u^{32*}} =({\la_d^{32*}}/{\la_u^{32*}})_{\rm cancel}(1-\delta)$.
Then we can take $\delta$ as a degree of required tuning for cancellation in {$\tau^- \to \mu^- \ga$}.
In Fig.~\ref{fig:h2mt_V}, the  black lines show a constant contour plot of $\delta$ in 
$(m_V,\lambda_{\rm conv} (\equiv |\lambda_u^{32} \lambda_e^{33}|))$-plane {in percentage terms} when we take the upper limit on $C_R^\gamma$ (with $C_L^\gamma = 0$) in Eq.~(\ref{eq:CRbound}).
{The plot shows that we need fine-tuning at the level of $0.1\%$ to explain the excess of $h \to \mu\tau$ consistently}.
}

\section{$h \to \mu \tau$}
\label{sec:h2mt}

\begin{figure}
\center
\includegraphics[width=13cm]{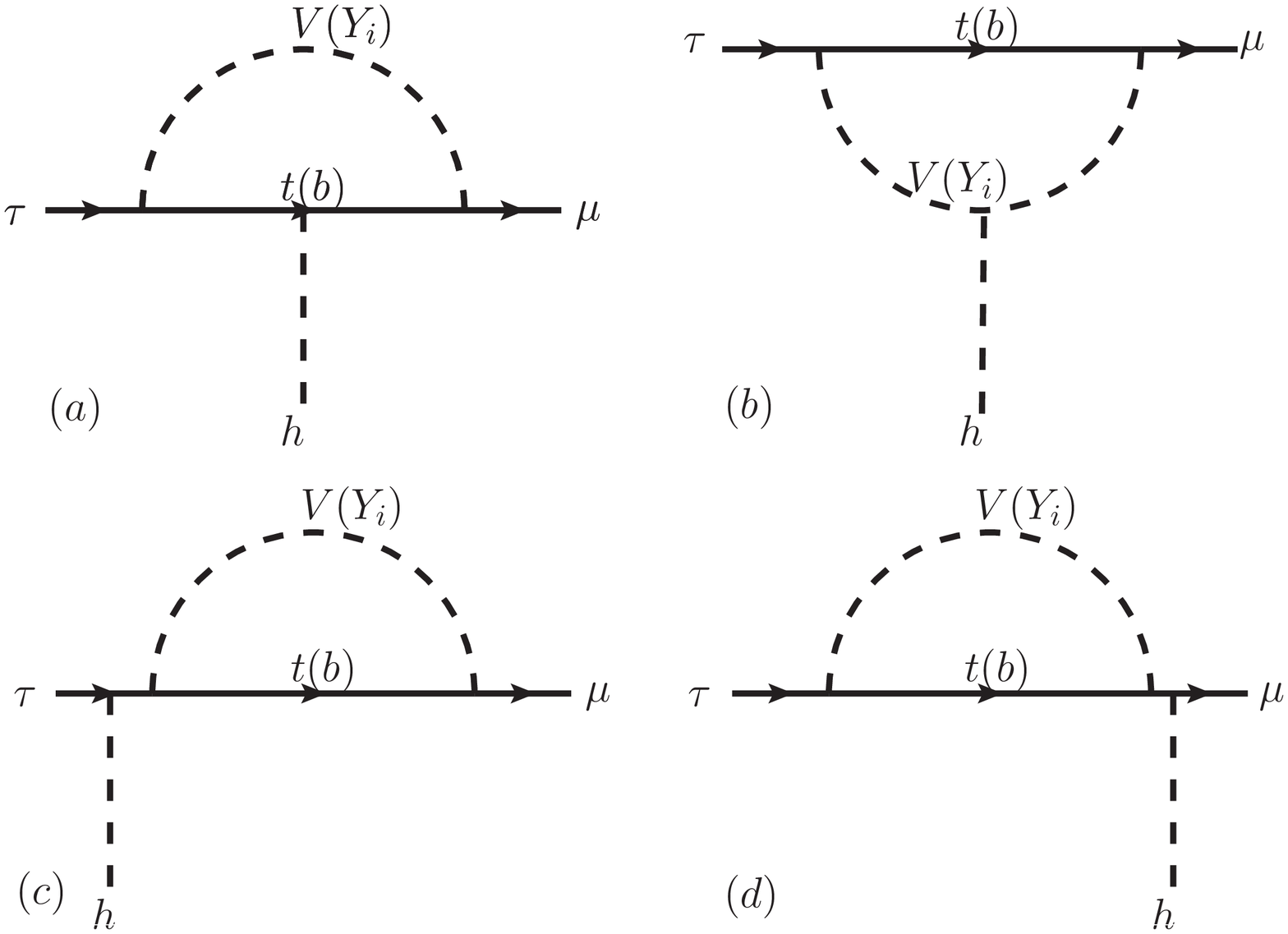}
\caption{Feynman diagrams of one-loop correction for $H^0-\mu-\tau$ vertex.}
\label{fig:h2mt}
\end{figure}

The lepton flavor violating Higgs decay is evaluated from the Feynman diagrams shown in Fig.~\ref{fig:h2mt}.
The divergence in diagram Fig.~\ref{fig:h2mt}\,(a) cancels those in Fig.~\ref{fig:h2mt}\,(c),\,(d), and the {total} result is finite, generating the dimension-four effective {operators}
\bea
{\cal H}_{\rm eff}(h \to \mu\tau) &=& h \ol{\mu} (C_R P_R + C_L P_L) \tau + {H.c}.
\label{eq:Wilson_htomutau}
\eea
The dimensionless effective couplings $C_{R,L}$ are calculated to be
\bea
C_R &=& - \lambda_u^{32*} \lambda_e^{33*} \frac{N_c m_t}{16 \pi^2 v} \Bigg[ I_{acd}(r_t,r_h) + I_a(r_t,r_h) - {\lambda_{HR}} \frac{v^2}{m_V^2} I_b(r_t,r_h) \Bigg] \nl
&& - \lambda_d^{32*} \lambda_e^{33*} \frac{N_c m_b}{16 \pi^2 v} \Bigg\{\sum_{i=1}^2 O_{2i} O_{1i} \Big[ I_{abc}(0,s_i) + I_a(0,s_i) \Big]\nl
&& - \sum_{i,j=1}^2 O_{2i} O_{1j} ( O^T \Lambda O)_{ji} \frac{v^2}{m_{Y_b}^2} I_{b2}(s_{ij},s_j) {\Bigg\}},
\label{eq:h2mt}
\eea
\vspace{-6mm}
\bea
{C_L} &=& - {\lambda_e^{23} \lambda_u^{33}} \frac{N_c m_t}{16 \pi^2 v} \Bigg[ I_{acd}(r_t,r_h) + I_a(r_t,r_h) -\lambda_{HX} \frac{v^2}{m_V^2} I_b(r_t,r_h) \Bigg] \nl
&& - {\lambda_e^{23} \lambda_d^{33}} \frac{N_c m_b}{16 \pi^2 v} \Bigg\{\sum_{i=1}^2 O_{2i} O_{1i} \Big[ I_{abc}(0,s_i) + I_a(0,s_i) \Big]\nl
&& - \sum_{i,j=1}^2 O_{2i} O_{1j} ( O^T \Lambda O)_{ji} \frac{v^2}{m_{Y_b}^2} I_{b2}(s_{ij},s_j) {\Bigg\}},
\label{eq:h2mt_2}
\eea
with $r_t=m_t^2/m_V^2, r_h=m_h^2/m_V^2$, $s_i = m_h^2/m_{Y_i}^2$, $s_{ij} = m_{Y_i}^2/m_{Y_j}^2$, and
\bea
\Lambda \equiv 
\left(\begin{array}{cc}
\lambda_{HR} & \lambda_{\rm mix} \\
\lambda_{\rm mix} & \widetilde{\lambda}_{HR}\\
\end{array}
\right).
\eea
Note that the coupling combinations, $\lambda_u^{32*} \lambda_e^{33*}$ and $\lambda_d^{32*} \lambda_e^{33*}$ in $C_R$; $\lambda_e^{23} \lambda_u^{33}$ and $\lambda_e^{23} \lambda_d^{33}$ in $C_L$, are also found in the terms in $C_R^\gamma$ and $C_L^\gamma$ for describing primary contributions to $\tau^{-} \to \mu^{-} \gamma$, respectively.
But here, {no sizable cancellation emerges} between terms being proportional to $m_t$ and $m_b$ when we adjust parameters for realizing the cancellation in $\tau^{-} \to \mu^{-} \gamma$.
We ignore the apparently irrelevant terms being proportional to $m_\tau$ or $m_\mu$, which arise from chirality flips in the external lines.
The loop functions are
\bea
I_{acd}(r_t, r_h) &=& -\frac{1}{2} - 2 \int [d x] \log\Big[x_3 + (1-x_3) r_t -x_1 x_2 r_h -i\varepsilon \Big] \nl
 && + \int_0^1 d x \log\Big[x + (1-x) r_t -i\varepsilon \Big], \nl
I_a(r_t,r_h) &=& \int [d x] \frac{x_1 x_2 r_h -r_t}{x_3 +(1-x_3) r_t - x_1 x_2 r_h}, \nl
I_b(r_t,r_h) &=& \int [d x] \frac{1}{1-x_3 +x_3 r_t - x_1 x_2 r_h}, \nl
I_{b2}(s_{ij},s_j) &=& \int [d x] \frac{1}{x_1 s_{ij} +x_2 - x_1 x_2 s_j}, 
\eea
where $ \int [d x] \equiv\int_0^1 d x_1 \int_0^1 d x_2 \int_0^1 d x_3 \, \delta(1-x_1-x_2-x_3)$ and $\varepsilon$ represents an infinitesimal positive value.\footnote{When $X$ is real, the relation $\log\left[ X \pm i \varepsilon \right] = \log\left[\left|X\right|\right] \pm i \pi \theta\left( -X \right)$ with the Heaviside {theta} function $\theta$ is useful.}
The form of the partial width $\Gamma_{h \to \mu^{-} \tau^{+}}$ is described by use of the Wilson coefficient $C_R$ and $C_L$ in Eq.~(\ref{eq:Wilson_htomutau}) as
\bea
\Gamma_{h \to \mu^{-} \tau^{+}} = \frac{\bar{\beta}}{16 \pi m_h}
\left[ {(m_h^2 - m_\mu^2 - m_\tau^2)} \left( |C_R|^2 + |C_L|^2 \right) - 2 m_{\mu} m_{\tau} \left( C_R C_L^\ast + C_L C_R^\ast \right) \right],
\eea
with the kinetic factor
\bea
\bar{\beta} = \sqrt{1 - \frac{2(m_\mu^2 + m_\tau^2)}{m_h^2} + \frac{(m_\mu^2 - m_\tau^2)^2}{m_h^4}},
\eea
while that of the conjugated process $\Gamma_{h \to \mu^{+} \tau^{-}}$ is straightforwardly obtained by the replacements $C_{R} \to C_{R}^\ast$ and $C_{L} \to C_{L}^\ast$.
The inclusive width $\Gamma_{h \to \mu \tau}$ is simply defined as
\bea
\Gamma_{h \to \mu \tau} = \Gamma_{h \to \mu^{-} \tau^{+}} + \Gamma_{h \to \mu^{+} \tau^{-}}.
\eea
{We use the value $\Gamma^{\text{SM}}_{h} = 4.07\,\text{MeV}$ in $m_h = 125\,\text{GeV}$ reported by the LHC Higgs Cross Section Working Group~\cite{Heinemeyer:2013tqa} for evaluating $\mathcal{B}(h \to \mu \tau)$ in our model.}

In the following analysis, as we did in the $\tau^{-} \to \mu^{-} \gamma$ in {Sec.}~\ref{sec:tau2mug}, we adopt the assumption of $C_L = 0$.
Among many terms in (\ref{eq:h2mt}), the two terms in the first line, {\it i.e.}, the top-quark contribution in Fig.~\ref{fig:h2mt}\,(a) dominates {and we ignore the bottom-quark contributions in the following numerical {estimation}}.
In Fig.~\ref{fig:h2mt_V}, we show the $2\sigma$ range to explain {the excess in $h \to \mu \tau$ shown in Eq.~(\ref{eq:CMS_h}) in} $(m_V, \lambda_{\rm conv})$-plane, where
$\lambda_{\rm conv} \equiv |\lambda_u^{32} \lambda_e^{33}|$.
{We set ${\lambda_{HR}} = 1$, which is the coupling of the subleading term in Eq.~(\ref{eq:h2mt}) with the suppression factor $v^2/m_V^2$.}
{Here, an upper limit on $\lambda_{\text{conv}}$ is estimated as $(\lambda_u^{32} \lambda_e^{33})|_{\text{max}} = (
  (\lambda_u^{32}/\lambda_d^{32}) \lambda_d^{32} \lambda_e^{33})|_{\text{max}} \simeq 0.25 \cdot 4\pi \cdot 4\pi \simeq
  40$, where $0.25$ means a typical maximal value of the ratio $(\lambda_u^{32}/\lambda_d^{32})$ shown in
  Fig.~\ref{fig:cancel} (when $m_V$ is multi TeV and $\sin{\alpha_Y} = 1/\sqrt{2}$), and $4\pi$ comes from perturbative
  regime in $\lambda_d^{32}$ and $\lambda_e^{33}$.}

Combining Fig.~\ref{fig:cancel} and Fig.~\ref{fig:h2mt_V}, we can see that it is possible to explain the excess shown in Eq.~(\ref{eq:CMS_h}) in our scenario.
At first, we will remember the relation in the LQ's masses, $m_{Y_i} \sim m_V/6$ for ensuring natural cancellations between $\lambda^{32}_{u}$ and $\lambda^{32}_{d}$ in $\tau^{-} \to \mu^{-} \gamma$.
When we request the (exact) cancellation in $\tau^{-} \to \mu^{-} \gamma$, as shown in Fig.~\ref{fig:cancel}, the ratio $(\lambda_u^{32}/\lambda_d^{32})^{\ast}$ should be smaller than unity.
Considering a typical scale of $m_V$ is more than a few TeV through the relation $m_{Y_i} \sim m_V/6$ and the latest LHC bounds on $m_{\text{LQ}}$, as a rough estimation, $\lambda_{\text{conv}}$ needs to be larger than {around} ten.
Taking into account the bound via perturbativity $\lambda_e^{33} < 4\pi$, roughly speaking, {$\lambda_{u}^{32}$} should be greater than one {through the definition of $\lambda_{\text{conv}}$}.
{Following this property}, we should think about the property of the ratio $(\lambda_u^{32}/\lambda_d^{32})^{\ast}$.
Roughly, greater than $0.1$ is required for realizing the {above} inequality ${\lambda_{u}^{32}} > 1$ within the region where $\lambda_d^{32}$ is still perturbative {($\lambda_d^{32} < 4\pi$)}.
This means that the mixing angle $\alpha_Y$ should be large to some extent since when $\alpha_Y$ becomes far from the maximal case, the region with $(\lambda_u^{32}/\lambda_d^{32})^{\ast} > 0.1$ shrinks or disappears.\footnote{{It is possible to modify the mass relation $m_{Y_i} \sim m_V/6$ without caring about the difference between $\lambda_u^{32}$ and $\lambda_d^{32}$.
When $m_V$ is heavier than the case following $m_{Y_i} \sim m_V/6$, the top contribution in $\tau^{-} \to \mu^{-} \gamma$ decreases and the ratio $(\lambda_u^{32}/\lambda_d^{32})^{\ast}$ can get to be large, which means that larger ${\lambda_u^{32}}$ would be realizable.
On the other hand, however, a large $m_V$ suppresses the process $h \to \mu \tau$.}}

As an example, we can satisfy $\tau^- \to \mu^- \gamma$ constraint, {with $(m_V, m_{Y_1}, m_{Y_2}) = (3.6, 0.9, 0.6)\,\text{TeV}$ and $\sin{\alpha_Y} = 1/\sqrt{2}$ leading to
$\lambda_u^{32}/\lambda_d^{32} \approx 0.15$. If we take $\lambda_d^{32} \approx 10$ and {$\lambda_e^{33} \approx 4$, we
get $\lambda_{\rm conv} \approx 6$}, which can explain the central value shown in Eq.~(\ref{eq:CMS_h}).}

\section{$(g-2)_\mu$}
\label{sec:g-2}

The anomalous magnetic moment of the muon has been measured to 0.5 ppm level~\cite{Bennett:2004pv},
\bea
 a_\mu^{\rm exp} = 116\,592\,080(63) \times 10^{-11}.
\eea
Theoretical calculation in the SM has similar precision~\cite{Miller:2007kk}
\bea
 a_\mu^{\rm SM} = 116\,591\,785(61) \times 10^{-11}.
\eea
The discrepancy
\bea
\Delta a_\mu = a_\mu^{\rm exp} - a_\mu^{\rm SM} = (295 \pm 88) \times 10^{-11}
\eea
is believed to come from new physics contributions.

{However, we should also keep in mind that there is a possibility that the discrepancy (or part of it) comes from underestimated uncertainties in hadronic part, for example, in hadronic light-by-light scattering.
Lattice calculations~{\cite{Blum:2014oka,Blum:2015gfa,Green:2015mva}} as well as calculations using dispersion relations~{\cite{Colangelo:2014dfa,Pauk:2014rfa,Colangelo:2015ama}} will reduce the hadronic uncertainties in the future.}

In our model the leptoquark contribution to $(g-2)_\mu$ is given by
\bea
\Delta a_\mu &=& -\frac{N_c m_\mu}{8 \pi^2 m_V^2}\Bigg[ 
m_\mu \Big(\left|\la_e^{23} \right|^2  +\left| \la_u^{32} \right|^2  \Big)
\Big(\frac{2}{3} I_1(x) +\frac{5}{3} J_1(x) \Big) 
+ {\rm Re}(\la_u^{32}  \la_e^{23})  m_t \Big(\frac{2}{3} I_2(x) +\frac{5}{3} J_2(x) \Big) \Bigg] \nl
&-& \sum_{j=1,2} \frac{N_c m_\mu}{8 \pi^2 m_{Y_j}^2}\Bigg[ 
m_\mu \Big( \left|\la_e^{23} \right|^2 O_{1j}^2   +\left| \la_d^{32} \right|^2 O_{2j}^2   \Big)
\Big(-\frac{1}{3} I_1(y_j) +\frac{2}{3} J_1(y_j) \Big) \nl
&+& {\rm Re}(\la_d^{32}  \la_e^{23})  O_{1j} O_{2j} m_b \Big(-\frac{1}{3} I_2(y_j) +\frac{2}{3} J_2(y_j) \Big) \Bigg],
\label{eq:g-2}
\eea
the loop functions are given in (\ref{eq:loop_ftns}).
We notice that, if we set $m_{\mu,\tau} \to 0$ {in $C_R^\gamma$ in Eq.~(\ref{eq:C7g})} and {inside the square brackets in $\Delta a_\mu$ in Eq.~(\ref{eq:g-2})}, $\Delta a_\mu$ is exactly proportional to $C_R^\gamma$ {as}
\bea
\Delta a_\mu = -\frac{4 m_\mu}{e} \frac{\la_e^{23}}{\la_e^{33}} C_R^\gamma,
\eea
where we assumed all the couplings are real. If we use the current upper bound of $C_R^\gamma$ in (\ref{eq:CRbound}), we get
\bea
\Delta a_\mu \approx -(66.3 \times 10^{-11}) \frac{\la_e^{23}}{\la_e^{33}}.
\eea
Therefore we see that, if {$-7 \, \lesssim \, {\la_e^{23}}/{\la_e^{33}} \, \lesssim \, -2$}, we can explain the
muon $(g-2)_\mu$ {with $\pm2\sigma$ accuracy}.

\begin{figure}
\center
\includegraphics[width=5cm]{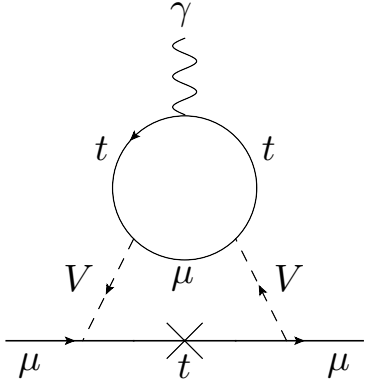}
\caption{A two-loop Barr-Zee type diagram for $(g-2)_\mu$.}
\label{fig:Barr-Zee}
\end{figure}

{Since in our case the Yukawa couplings to explain $h \to \mu\tau$ and $(g-2)_\mu$ are rather large, one may expect higher order diagrams such as Barr-Zee type two-loop diagrams~\cite{Barr:1990vd} may enhance $(g-2)_\mu$ as in the case of MSSM with large $\tan\beta$~\cite{Arhrib:2001xx,Baek:2004tm}.
In our estimate, the dominant two-loop diagram is shown in Fig.~\ref{fig:Barr-Zee}.
Other diagrams, such as the one with LQs running inside the loop, are suppressed, for example, by small muon mass,
and we do not consider them.
Although the diagram in Fig.~\ref{fig:Barr-Zee} may look comparable with the one-loop diagrams due to large $\lambda_u^{32}
\lambda_e^{23}$, we still need chirality flip inside the muon line in the fermionic triangle loop.
Concretely, the diagram is estimated to be suppressed at least by
\bea
\sim {1 \over 16 \pi^2} \lambda_u^{32} \lambda_e^{33} \frac{m_\mu}{m_t}  \sim 10^{-3}
\eea
compared to the one-loop diagram.}

\section{Conclusion}
\label{sec:conc}

In this paper, we considered the recent CMS excess in $h \to \mu\tau$ and the muon $(g-2)$ anomaly.
We showed that we can accommodate both discrepancies by introducing two leptoquarks 
$R_2(3,2,7/6)$ and $\wt{R}_2(3,2,1/6)$ 
{that are free from proton decay problems at renormalizable level}.
The constraints from lepton flavor violating process $\tau^- \to \mu^- \gamma$ can be evaded by {a natural} cancellation between leptoquark contributions with some tuning {on $\lambda_u^{32}$ and $\lambda_d^{32}$, where their orders can be the same.
When the cancellation is realized, sizable couplings contributing to $h \to \mu \tau$ are allowed and then we give a reasonable explanation on the excess.
The $(g-2)_{\mu}$ anomaly is also explained. 
Finally, we mention that various kinds of other anomalies in flavor physics have been reported~\cite{BaBar_tau,BaBar_tau13,LHCb_Kstar,LHCb,LHC_tau,Belle_tau}.
Giving a more exhaustive explanation in the context of leptoquarks would be an important task{~\cite{B-leptoquark}}.


\acknowledgments
{K.N.} is grateful to Tomohiro~Abe, {Shinya~Kanemura}, Kin-ya~Oda, Makoto~Sakamoto and {Koji~Tsumura} for fruitful discussions.
This work is supported in part by  National Research Foundation of Korea (NRF) Research Grant
{No.}~NRF-2015R1A2A1A05001869 ({S.B.}).

\end{document}